\documentclass[letterpaper, 10 pt, conference]{ieeeconf}  

\usepackage{graphics} 
\usepackage{xcolor}
\usepackage{soul}
\usepackage[utf8]{inputenc}
\usepackage{booktabs}
\usepackage{times}
\usepackage{graphicx}
\usepackage{tikz}
\usetikzlibrary{positioning}

\usepackage{amsmath,amsfonts,amssymb}
\usepackage{thmtools,thm-restate}	
\usepackage[caption=false]{subfig}
\allowdisplaybreaks	
\usepackage[linesnumbered,ruled,vlined]{algorithm2e} 
\newtheorem{theorem}{Theorem}[section]
\newtheorem{proposition}[theorem]{Proposition}

\newtheorem{definition}[theorem]{Definition}

\newcommand\Loss{\boldsymbol{\mathcal{\ell}}}

\newcommand\Path{\boldsymbol{p}}
\newcommand{\path}{p}
\newcommand\Pathset{\mathcal{P}}
\newcommand\Edgeset{\mathcal{E}}
\newcommand\Cban{\textsc{CBand}}
\newcommand\EdAl{\textsc{Edge}}
\newcommand\CB{\mathcal{CB}}
\newcommand\ComAl{\textsc{ComBand}}
\newcommand\muFree{\mu_{\textrm{free}}}
\newcommand\muUni{\mu_{\textrm{uni}}}
\title{\LARGE \bf
Combinatorial Bandits for Sequential Learning in Colonel Blotto Games
}

\author{ \parbox{2.1 in}{\centering  Dong Quan Vu\\
        AAAIRD Department\\
        Nokia Bell Labs, Paris Saclay, Nozay, France
        }
        \hspace*{ 0.08 in}
        \parbox{2.1 in}{ \centering Patrick Loiseau\\
        Univ. Grenoble Alpes, Inria, CNRS, Grenoble INP, LIG, France \& MPI-SWS, Germany
        }
        \hspace*{ 0.08 in}
        \parbox{2.1 in}{ \centering Alonso Silva\\
        Safran Tech, Signal and Information Technologies,\\ 
        Magny-Les-Hameaux, France
        }
}

\begin{document}
\maketitle
\thispagestyle{empty}
\pagestyle{empty}
\begin{abstract}
The Colonel Blotto game is a renowned resource allocation problem with a long-standing literature in game theory (almost 100 years). However, its scope of application is still restricted by the lack of studies on the incomplete-information situations where a learning model is needed. In this work, we propose and study a regret-minimization model where a learner repeatedly plays the Colonel Blotto game against several adversaries. At each stage, the learner distributes her budget of resources on a fixed number of battlefields to maximize the aggregate value of battlefields she wins; each battlefield being won if there is no adversary that has higher allocation. We focus on the bandit feedback setting. We first show that it can be modeled as a path planning problem. It is then possible to use the classical $\ComAl$ algorithm to guarantee a sub-linear regret in terms of time horizon, but this entails two fundamental challenges: (i) the computation is inefficient due to the huge size of the action set, and (ii) the standard exploration distribution leads to a loose guarantee in practice. To address the first, we construct a modified algorithm that can be efficiently implemented by applying a dynamic programming technique called weight pushing; for the second, we propose methods optimizing the exploration distribution to improve the regret bound. Finally, we implement our proposed algorithm and perform numerical experiments that show the regret improvement in practice.
\end{abstract}

\section{INTRODUCTION} 
The \textit{Colonel Blotto game} (henceforth, $\CB$ game) is a classical resource allocation problem where players simultaneously distribute a fixed number of troops (indivisible resources) on a certain number of battlefields to maximize the aggregate value of battlefields they win, each battlefield being won by the player who allocates the most resources to it. The $\CB$ game can be used to model a vast range of practical situations, e.g., allocating security forces in security (\cite{arce2012,chia2011}); allocating broadcasting time in advertisement (\cite{masucci2014}); allocating shared spectrum in wireless networks (\cite{hajimirsaadeghi2017dynamic}) and allocating resources to persuade voters in politics (\cite{hortala2012}). 

The $\CB$ game was first introduced by \cite{borel1921} in 1921 and has been extensively studied after, especially in recent years. In particular, the Nash equilibrium of the relaxed continuous version (where resource allocations can be fractional) was studied by \cite{grosswagner,roberson2006,schwartz2014} under different assumptions. For the discrete version (with indivisible troops), \cite{Behnezhad17a} proposed polynomial (but still expensive) algorithms to compute the exact equilibrium whereas \cite{Vu18a} proposed a much faster algorithm to compute an approximate equilibrium. All these works, however, only consider the full information one-shot game setting.\footnote{A few works studied dynamic settings of the game, but only limited to two or three stages and they focus on asynchronous allocations (\cite{gupta2014,powell2009}).} 
In most of the applications, the most natural setting is an incomplete information repeated game. For example, in advertising, one can consider an online marketing campaign which once per day reviews how the marketing campaign performs and based on this information, learns a better strategy. Resources and strategies of the adversaries are usually unknown or need to be estimated by the learner; especially in the dynamic setting. The question is then how to design a good sequential resource allocation~policy. 

The $\CB$ game has a specific combinatorial structure and the most natural way to model sequential learning in this game is to use the Combinatorial Bandit (henceforth, $\Cban$) framework, defined as follows: at each stage $t$ within a time horizon~$T$, the learner chooses a vector $\Path^t$ in her action set $S \subset \{0,1\}^E$, for an $E \in \mathbb{N}$; then a loss vector \mbox{$\Loss^t \in [0,1]^E$} is generated by the adversaries; the learner suffers a scalar loss \mbox{$L\left(\Path^t\right)=(\Loss^t)^{\top} \Path^t$}. The learner's objective is to minimize her \emph{expected regret}
$R_{T}$, i.e., the difference between her cumulative loss and that of her single best-action in hindsight, formally defined as~follows:
\begin{equation}
R_{T} := \mathbb{E}\left[ {\sum \limits_{t=1}^{T} L\left( \boldsymbol{\Path}_t\right)} - \min\limits_{\boldsymbol{\Path} \in S} {\sum \limits_{t=1}^{T}{L}\left( \boldsymbol{\Path}\right)} \right].
\label{RegretDef}
\end{equation}
Importantly, in $\Cban$, the feedback that the learner receives at the end of each stage is under the \emph{bandit setting}: the learner's only observation when stage $t$ ends is the scalar loss~\mbox{$L\left(\Path^t\right)$}. This is the most generic information-setting considered in the literature. This setting covers many applications of the $\CB$ game; e.g., in advertising, the total profit of selling a product can be easily observed while it is much harder to keep track of the partial profit of each ad (simultaneously promoting that~product).

$\ComAl$ algorithm, proposed by \cite{cesa2012}, is a classical algorithm for solving $\Cban$s. It provides a regret guarantee of the order $\mathcal{O}(\sqrt{TE\log{|S|}})$ that improves significantly than naively using other standard Multi-armed Bandit algorithms. In the $\CB$ game (modeled as a $\Cban$), $E$ is polynomial in the number of battlefields and~troops while $|S|$ is exponential in terms of these parameters. However, applying directly $\ComAl$, we face two important~challenges. 

\paragraph{Challenge 1: Computation Issue} The main problem with the $\ComAl$ algorithm is that in general, it cannot be implemented efficiently (its running time is in $\mathcal{O}(|S|T)$). However, in some special cases, there are techniques that allow us to efficiently implement variants of this algorithm. The \textit{path planning problem} (henceforth, PPP) is such an example: each action is equivalent to a path on a graph and the loss of a chosen path equals to the aggregation of losses of edges on that path. In PPPs, \cite{sakaue2018} recently proposed an efficient variant of $\ComAl$ running in $\mathcal{O}(E^2 T)$ based on the weight-pushing technique (introduced by \cite{takimoto2003}). However, this algorithm has a redundancy in representation (involving 5 sub-algorithms) and is still non-trivial to be implemented. A direct application of $\ComAl$ to the $\CB$ model is impractical; can we find a new representation of $\CB$ game to obtain a path planning model allowing an efficient implementation of $\ComAl$? If the answer is positive, we also desire a simpler representation of this efficient algorithm.  

\paragraph{Challenge 2: Optimizing Exploration Distribution} $\ComAl$ mixes an exploitation procedure (updated according to an unbiased loss estimator) with an \textit{``exploration distribution"} on the action set. 
The regret bound given by $\ComAl$ algorithm depends directly on the choice of an exploration distribution to be~used. For PPPs (and also for $\CB$ games), the optimal exploration distribution remains~unknown (this is an open question proposed~by~\cite{cesa2012}).


\emph{Our Contributions:} In this paper, we provide the first analysis of sequential learning in $\CB$ games. The action set of the $\CB$ game can be represented by a special graph; thus, we can model it as a PPP. Focusing on this model, our contribution is twofold: ${(i)}$ Based on the weight pushing technique, we construct a simple algorithm, called $\EdAl(\mu)$, that can be efficiently implemented and it guarantees a polynomial regret bound in terms of the $\CB$ game's parameters; ${(ii)}$ We propose a fast method to compute an exploration distribution that can be used as the input of $\EdAl(\mu)$ to improve the regret bound. Numerical experiments are conducted to illustrate this improvement, both in terms of the performance and the computation time.

Although in this work, we focus only on the $\CB$ game, note that our setting is more general. Our results can be extended to PPPs with general graphs that includes many other resources allocation games and multi-task online optimization. Note finally that \cite{combes2015combinatorial} proposed another variant of $\ComAl$ (mixing with ideas of the OSMD algorithm proposed by \cite{AudibertBL2014}), called the \textsc{CombEXP} algorithm, that improves the complexity of $\ComAl$ in several cases while maintaining the regret guarantees. However, PPP is not explicitly considered in~\cite{combes2015combinatorial} and it remains an open question whether any arbitrary instance of the PPP satisfies the condition such that \textsc{CombEXP} can be efficiently implemented (i.e., the convex hull of the action set can be represented by a polynomial number of linear equations and linear inequalities). Therefore, $\ComAl$ is still the state-of-the-art algorithm for our considering problem. Moreover, \textsc{CombEXP} also uses the uniform exploration distribution that is sub-optimal in PPPs (see also \cite{cesa2012}); thus, our second contribution in finding better exploration distributions is~relevant.

\emph{Notation:} Throughout the paper, we use bold symbols (e.g., $\boldsymbol{x}$) to denote (column) vectors with subscript indexes (e.g., $x_i$) to denote its elements. On the other hand, the superscript~$t$ refers to the stage and the notation $[k]$ denotes the set $\{1,2,\ldots,k\}$, for any $k \in \mathbb{N} \setminus\{0\}$. In graphs, we use the notation $e \in \Path$ to refer that the edge $e$ belongs to the path $\Path$. Finally, $\top$ denotes the transpose matrix/vector and $\mathbb{M}_{k \times k'}$ is the set of all real matrices with dimension $k \times k'$.
\section{PRELIMINARIES}
\label{sec:Comband}
In this section, we review the standard $\ComAl$ algorithm (proposed by~\cite{cesa2012}). We also highlight its drawbacks that need be~improved. A pseudo-code of $\ComAl$, written in our notation, is given as Algorithm~\ref{AlgoBianchi}.
    \begin{algorithm}[h!]
    \DontPrintSemicolon
     \KwIn{$S\!\subset\! \{0,1\}^E$, $T\! \in \! \mathbb{N}$, $\gamma \!\in\! [0,1], \eta\!>\!0$, distribution $\mu$ on $S$.}
      $\forall \Path \in S $, $w^{1}(\Path):= 1$.\;
     \For{$t=1,2,\ldots,T$}{
     	Adversaries choose the loss vector $\Loss^t$ (unobserved by the learner). \;
      	$\forall \Path \in S$, $\nu^t(\Path):={w_t({\Path})} / {\sum_{\boldsymbol{q} \in S} {[w_t(\boldsymbol{q})]}}$.\;
        Sample and play ${\Path}^t$ according to \mbox{$d^t(\Path)\!=\! (1\!-\!\gamma)\nu^t(\Path)\! +\! \gamma \mu(\Path)$}.\;
        Suffer and observe the loss $L(\Path^t)\!=\!{\left(\Loss^t\right)\!}^{\top}\!{{\Path}}^t \le 1$.\;
        Compute $C^t:= \mathbb{E}_{\Path \sim d^t}[\Path \Path^{\top}] \in \mathbb{M}_{E \times E}$ .\;    
        Compute the estimated loss $\hat{\Loss}^t := L(\Path^t) \left(C^{-1}_t {\Path}^t \right) = \left(\Loss^t ({{\Path}^t})^{\top} \right) C^{-1}_t {\Path}^t$. \;
        $\forall \Path \in S$, $w^{t+1}(\Path):= w^t(\Path)e^{-\eta (\hat{\Loss}^t)^{\top}{\Path} }$.
        }
     \caption{$\ComAl$($\mu$) for $\Cban$.} \label{AlgoBianchi}
    \end{algorithm}

%
At each stage~$t$, $\ComAl$ keeps a weight $w^t(\Path)$ for each action $\Path$ and it samples an action (line 5) from a distribution, called $d^t$, mixing between an ``exploitation" distribution $\nu^t$ (normalization of the action weights) and an ``exploration" distribution $\mu$ (unchanged over time). An unbiased estimator $\hat{\Loss}^t \in [0,1]^E$, based on the ``co-occurrence" matrix \mbox{$C^t:=  \mathbb{E}_{\Path \sim d^t}[\Path \Path^{\top}] \in \mathbb{M}_{E \times E}$}, is used to estimate the loss vector $\Loss^t$. Then, the action weights are updated by the exponential rule using these estimated losses (line 9).

In $\ComAl$, the exploration distribution $\mu$ is chosen a priori and it can be any arbitrary distribution on $S$ such that $S$ is spanned by the support of $\mu$. Importantly, the performance guarantee of $\ComAl$ depends directly on the choice of $\mu$; to highlight this, we henceforth parameterize $\ComAl$ with $\mu$ and use the notation $\ComAl$($\mu$). Consider the matrix \mbox{${M}(\mu) = \mathbb{E}_{\Path \sim \mu} {\left[ \Path \Path^{\top} \right]}$}, we denote by $\lambda^*[M(\mu)]$ the \textit{smallest nonzero eigenvalue} of ${M}(\mu)$ and let \mbox{$n:=\max\{ \|\Path \|_{1}, \Path \in S  \}$}. An upper-bound of the expected regret given by this algorithm is stated as follows.
\begin{theorem}
\label{BianchiTheo}
\emph{In any~$\Cban$ problem,~the $\ComAl(\mu)$ algorithm with appropriate parameters yields an expected regret \mbox{$R_T \le 2 \sqrt{\left[ {2n}/({E \cdot \lambda^*[M(\mu))}] \!+\!1\right] T E \log(|S|)}$}.}
\end{theorem}

This theorem is extracted from Theorem 1 in \cite{cesa2012} and is rewritten here under our notation. Trivially, the larger $\lambda^*[M(\mu)]$ is, the better the regret bound that $\ComAl(\mu)$ guarantees. The problem of optimizing $\mu$ and $\lambda^*[M(\mu)]$ in general $\Cban$s (and particularly for PPPs) remains an open question (see \cite{cesa2012} for several positive~examples).
Regarding the computation complexity of $\ComAl$, given a time horizon $T$, it runs in $\mathcal{O}(|S| \cdot T)$. Since $|S|$ is exponential in terms of $E$, it is inefficient to implement $\ComAl$. This is due to the \emph{weights-updating step} (line 9), the \emph{sampling step} (line 5) and the computation of the \emph{co-occurrence matrix} (line 7). We will analyze these steps in Section~\ref{Dyntech} and provide alternative procedures to improve them in the sequential learning model of Colonel Blotto games.

\section{COMBINATORIAL BANDIT MODEL OF LEARNING IN COLONEL BLOTTO GAMES}
\label{sec:Model}
We consider a sequential learning problem that involves a \textit{learner}, $A$~adversaries, $n$~battlefields and a time horizon~$T$ ($n \ge 2$ and $T>0$ are known by the learner, $A\ge 1$). Each battlefield $i~\in~[n]$ has a fixed value $b_i>0$ (hidden from the learner) and we assume normalized values, that is $\sum_{i=1}^n{b_i} = 1$. At each stage $t \in [T]$, the learner faces a decision problem of distributing $m$ troops ($m\ge 1$ is fixed) towards the battlefields while the adversaries simultaneously allocate theirs. The learner's allocations have to satisfy the budget constraint, that is she chooses a strategy $\Path^t$ in the action set \mbox{$S:=\{ \Path \in \mathbb{N}^n: \sum_{i=1}^n \nolimits{\path_i} = m\}$}.\footnote{The constraint \mbox{$\sum_{i=1}^n \nolimits{p_i}\!\le\!m$} is sometimes considered in the literature. However, since unallocated troops do not contribute to the payoff, the learner always has incentives to use all her troops.}
For any $i \in [n]$, the element $\path_i$ of strategy $\Path$ represents the quantity of troops she allocates to battlefield $i$. At the end of time~$t$, the learner suffers a loss $L\left( \Path^t \right)$ equal to the sum of values of battlefields that she loses, i.e., where there is at least one adversary having strictly higher local allocation than her. Without loss of generality, in case there are players who have tie allocations which are the highest in battlefield $i$, the value $b_i$ is shared equally among them. When $t$ ends, the learner observes the scalar number $L(\Path^t)$ but she does not know which battlefield she lost or won nor the strategies that the adversaries used (the bandit feedback). Note that the incurred loss is bounded, i.e., $L(\Path^t) \le 1, \forall \Path^t, \forall t$. The cumulative loss $\sum_{t=1}^T \nolimits {L(\Path^t)}$ is computed and the learner's objective is to minimize her expected regret (defined in~\eqref{RegretDef}). Henceforth, we refer to this model as $\CB(m,n)$~problem.

It is important to note that the size of the learner's strategy set is \mbox{$|S|=  \binom{n+m-1}{n-1} = \mathcal{O}(2^{\min\{n-1,m\}})$}, that is exponential in terms of $m$ (or $n$). Our objective is to design an algorithm guaranteeing an expected regret that is sub-linear in $T$ and polynomial in terms of $m$ and $n$ while its complexity is also polynomial in $m,n$, and~$T$. 
\subsection{Layered Graph and Learning in $\CB$ Games as PPPs}
\begin{figure}[htb!]%
\centering
	   \begin{minipage}{4.3cm}
\includegraphics[height=0.14 \textheight]{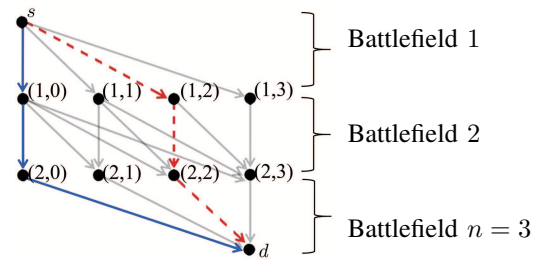}
\end{minipage}
\begin{minipage}{2.7cm}
Battlefield $1$ \\
\\ \\ 
Battlefield $2$\\
\\ \\ 
Battlefield $n=3$
\end{minipage}
    \caption{The graph $G_{3,3}$ corresponding to the game with $m=n=3$. Each path from $s$ to $d$ represents a strategy in $S$; e.g., the bold-blue path represents the strategy~$(0,0,3)$ while the dash-red path represents the strategy~$(2,0,1)$.}
    \label{fig1}
\end{figure}

In this section, we restate the formulation of $\CB(m,n)$ as a PPP that allows an efficient implementation of~$\ComAl$. To do this, for each $\CB(m, n)$ game, we design a special directed acyclic graph, called $G_{m,n}$, such that there exists a one-to-one mapping between the action set $S$ and the set of all paths of $G_{m,n}$; we call this the \emph{layered graph}.\footnote{This term is inspired by a graph proposed by~\cite{Behnezhad17a} that looks similar to $G_{m,n}$; however, it is used for a completely different purpose and it also contains more edges and paths than $G_{m,n}$ (that are not useful in this work).} The illustration of an instance of such a graph is presented in Figure~\ref{fig1}. The proof of existence of $G_{m,n}$ can be intuitively seen in Figure~\ref{fig1} and a formal definition is as follows (note that this definition is extracted from~\cite{vu2019colonel}). 

\begin{definition}[Layered Graph]
\label{CB_graph}
The graph $G_{m,n}$ is a DAG that~contains:
\begin{itemize}
\item[$(i)$] \mbox{$N:=2\!+\!(m\!+\!1)(n\!-\!1)$} nodes that are arranged into \mbox{$n\!+\!1$} layers. Layer $0$ has only one node \mbox{$s\!:=\!(0,0)$}, called the source node and Layer $n$ contains one node \mbox{$d\!:=\!(n,m)$}, called the destination node. Each Layer \mbox{$i \in [n-1]$} contains \mbox{$m+1$} nodes whose labels are ordered from left to right by \mbox{$(i,0),(i,1),\ldots,(i,m)$}.
\item[$(ii)$] There are directed edges from the source node $s$ to all nodes in Layer $1$ and from all nodes in Layer $n-1$ to the destination node $d$. For any Layer \mbox{$i \in [n-2]$}, there is a directed edge from node $(i,j_1)$ to node \mbox{$(i+1,j_2)$} (of Layer $(i+1)$) if $0 \le j_1 \le j_2 \le m$.
\end{itemize}
\end{definition}

%
%
%

Let $\mathcal{N}$ denote the set containing all nodes of $G_{m,n}$ (including the source and destination nodes). $G_{m,n}$ has \mbox{$E=\! (m\!+\!1)\left[4\! +\! (n\!-\!2)(\!m\!+\!2) \right]\!/2 = \mathcal{O}(n m^2)$} (directed) edges that are arranged into $n$ layers. Each edge represents an allocation of a certain number of troops toward a certain battlefield: the edge from node $(i,j_1)$ to node $(i+1,j_2)$ represents the allocation of player that puts $(j_2 - j_1)$ troops to battlefield $i+1$, for any \mbox{$i \in \{0,\ldots,n\!-\!1\}$}; for instance, in Figure~\ref{fig1}, the edge from $(1,0)$ to $(2,3)$ represents allocating 3 troops to Battlefield~2. We denotes $\Edgeset$ the set containing all the edges. Hereinafter, referring to the layered graph, we simplify the notations and use the term ``paths" to indicate the paths starting from $s$ and ending at $d$ if there is no other explicit explanation.  We define the set $\mathcal{P} \subset \{0,1\}^E$ containing all such paths and \mbox{$P :=  |\Pathset| = |S|=  \binom{n+m-1}{n-1} = \mathcal{O}(2^{\min\{n-1,m\}})$}. Given a strategy \mbox{$\Path \in S \subset[0,m]^n$}, we slightly abuse the notation and re-use $\Path= (\path_e)_{e\in \Edgeset}$ to denote the $E$-dimension $0$-$1$-vector that represents the path equivalent to this strategy. Particularly, $\forall e \in \Edgeset$, $\path_e = 1$ if and only if edge $e$ belongs to path~$\Path$ and $\path_e =0$ otherwise.

Finally, for the sake of completeness, we write down formally the PPP that is equivalent to the $\CB(m,n)$ model. At each stage $t$, the allocations of the learner and the adversaries to the battlefields determine a scalar loss
\mbox{$\Loss^t_e \in [0,1]$} (following the rule of the $\CB$ game) embedded on each edge $e \in \Edgeset$ of the graph~$G_{m,n}$. The learner has to choose a path \mbox{$\Path^t \in \mathcal{P}\subset \{0,1\}^E$} in $G_{m,n}$. The learner then suffers a loss \mbox{$L(\Path^t)= {\left(\Loss^t\right)}^{\top} {\Path}^t =\sum_{e \in \Path^t}\nolimits{\Loss^t_e}$} which equals the sum of losses from all edges belonging to the chosen path $\Path^t$. At the end of stage $t$, the learner only observes the scalar loss $L(\Path^t)$ of her chosen path but she does not know the loss of each edge. Henceforth, we focus our analysis on this model and we refer to it as \textsc{PathCB}$(m,n)$ to distinguish with $\CB(m,n)$.

\section{EFFICIENT ALGORITHM FOR PATH PLANNING PROBLEMS}
\label{Dyntech}

In this section, we revisit a standard dynamic programming technique, called weight pushing. This technique is the basic for the efficient implementation of $\ComAl$ algorithm in PPPs. The first idea of weight pushing technique could be tracked back to \cite{gyorgy2007} and \cite{takimoto2003} although it was only applied to efficiently sample a path in PPPs according to updating-rules based on the weights of edges. 
Recently, \cite{sakaue2018} proposed an application of weight pushing to compute the co-occurrence matrix $C^t$ in polynomial time in $E$; particularly for $\ComAl$ in PPPs with the Zero-suppressed Binary Decision Diagrams. This computation requires 5 sub-algorithms that involves heavy notations and unneccessary complexity for our setting. We restate the technique with the notations and parameters of \textsc{PathCB}$(m,n)$ and propose a more computationally efficient version of $\ComAl$.

\subsection{Paths and Edges' Weights}
\label{edges}

In PPPs, we call the action weights $w^t(\Path)$ involved in $\ComAl$ as the \emph{path weights} and recall that \mbox{$w^{t+1}(\Path):= w^t(\Path) e^{-\eta( \hat{\Loss}^t)^{\top} \Path }$}. At stage $t$, for each edge $e \in \Edgeset$, we define the \textit{edge weight} $w^t_e$ such that \mbox{$w^1_e := 1, \forall e$} (by convention) and \mbox{$w^{t+1}_{e}:= w^t_e \cdot e^{-\eta (\hat{\Loss}_e^t)}$}. It is trivial to deduce that \mbox{$w^{t}(\Path) = \prod_{e\in \Path}\nolimits{w^{t}_e}$}, $\forall \Path \in \Pathset, t\in [T]$, i.e., the weight of a path is the product of weights of all edges belong to it. The basic idea of weight pushing is to keep track of the paths weights (there is an exponential number of them) via the edges weights (only a polynomial number of them) by exploiting the structure of the graph.

Now, let us denote by $\Pathset_{(u,v)}$ the set of all paths starting from node $u$ and ending at node~$v$. Then, for each pair of nodes $(u,v) \in \mathcal{N} \times \mathcal{N}$, at stage~$t$, we~define \mbox{$H^t(u,v) := \sum \nolimits_{\Path \in \Pathset_{(u,v)}}{\prod \nolimits_{e \in \Path}{w^t_e} }$}. Intuitively, $H^t(u,v)$ is the sum of weights of all paths from node $u$ to node~$v$. Importantly, by conventionally setting $H^t(u,u):= 1$, \mbox{$\forall u \in \mathcal{N}$} and $H^t(u,v) =0$ if $\Pathset_{(u,v)} = \emptyset$, we can compute all the quantities $H^t(u,v)$ in $\mathcal{O}(E)$ time via the following procedure (see also \cite{gyorgy2007}): we first re-label the nodes set by \mbox{$\mathcal{N}\!=\! \{s=u_0,u_1,\ldots, d\!=\!u_N\}$} such that if there exists an edge connecting $u_i$ to $u_j$ then $i<j$. Then, for any $v\in \mathcal{N}$, recursively for \mbox{$ u \in \{ v-1,v-2, \ldots, s:=0\}$}, we can compute $H^t(u,v)$. 

%
\subsection{Sampling by Edges' Weights}
Inspired by Theorem 3 in \cite{gyorgy2007}, we can design an algorithm, denoted WP Algorithm (WP stands for weight pushing), that takes $w^t(e), e \in \Edgeset$ as inputs and outputs a path in $\Pathset$. More importantly, the probability that a path $\Path$ is an output of the WP Algorithm at stage $t$ is exactly $\nu^t(\Path)$---the exploitation distribution defined in~$\ComAl$. We rewrite this algorithm under our notation as Algorithm~\ref{Sample}. In Algorithm~\ref{Sample}, we denote by $e_{[u,v]}$ the edge connecting from node $u$ to node $v$ and by \mbox{$\mathcal{C}(u):=\{u^{\prime}>u: e_{[u,u^{\prime}]} \in \Edgeset\}$} the set of all direct children of $u$. 

\begin{algorithm}[htb!]
    \DontPrintSemicolon
     \KwIn{$G_{m,n}$, $t \in [T]$, $w^t_e, \forall e \in \Edgeset$.}
      Initialize $\mathsf{Q} := \{0\}$, $u_0=s$ and $k=0$.\;
      \For{$k \le n$}{
    Sample a node $u_{k+1}$ from $\mathcal{C}(u_k)$ with probability $w^t_{e_{[u_k, u_{k+1}]}} {H^t(u_{k+1},d)} \big/ {H^t(u_k,d)}$.\;
    Add $u_{k+1}$ to the set $\mathsf{Q}$.
    }
    \KwOut{$\Path^t \in \Pathset$ going through all nodes in~$\mathsf{P}$.}
     \caption{WP Algorithm: Sampling by edges'~weights.} \label{Sample}
\end{algorithm}
\subsection{Co-occurrence Matrix Computation}
We now turn our focus to the matrix \mbox{$C^t:=  \mathop{\mathbb{E}}_{\Path \sim d^t}[\Path \Path^{\top}]$} needed to be computed at each stage $t$ in the $\ComAl(\mu)$ algorithm. A direct computation of this matrix involves a sum of $P$ terms, that leads to the inefficiency of $\ComAl(\mu)$.  We first consider the following assumption on $\mu$:
 
\emph{Assumption 1: There exists a set of edges weights \mbox{$\tilde{\boldsymbol{w}}_e >0$}, $e\in \Edgeset$ such that for each path $\Path^* \in \Pathset$, we have \mbox{$\mu(\Path^*) = \prod \nolimits_{e\in \Path^*}{\tilde{w}_e} / \sum \nolimits_{\Path \in \Pathset}{(\prod \nolimits_{e\in \Path}{\tilde{w}_e})}$.}}

Intuitively, if $\mu$ satisfies Assumption 1, there exists a set of edges weights such that each path weight (according to $\mu$) equals to the multiplication of the weights of the corresponding edges. Note that the uniform distribution on $\Pathset$ (used by most of works in the literature) satisfies Assumption~1. 

Now, we observe that each entry ${C^t}_{e_1, e_2}$ equals to the probability that a chosen path $\Path^t \sim d^t$ contains both edges $e_1$ and $e_2$ (hence the name co-occurrence matrix). Formally, we have \mbox{$ C^t_{e_1, e_2} = \sum_{\Path \in \Pathset} d^t(\Path){\Path_{e_1} \Path_{e_2}} = \sum_{\{\Path: e_1, e_2 \in \Path\}} d^t(\Path)$}. Now, we define \mbox{$M(\nu^t):={\mathbb{E}}_{\Path \sim \nu^{t}(\Path)}[\Path \Path^{\top}]$} and \mbox{$M(\mu)={\mathbb{E}}_{\Path \sim \mu(\Path)}[\Path \Path^{\top}]$}---the co-occurrence matrices corresponding to distribution $\nu^t$ and $\mu$, respectively. From the definition of $d^t(\Path)$ in $\ComAl(\mu)$, we can~rewrite 
\begin{equation}
    C^t = (1-\gamma)M(\nu^t) + \gamma M(\mu) \label{eq:MatC}. 
\end{equation}
Therefore, to efficiently compute $C^t$, we need to efficiently compute $M(\nu^t)$ and $M(\mu)$. We do this by designing an algorithm, called Algorithm~\ref{Matrix}, based on the quantities $H^t(u,v)$ computed in the previous section.  Algorithm~\ref{Matrix} runs in $\mathcal{O}(E^2)$ time. $M(\nu^t)$ can always be computed by Algorithm~\ref{Matrix} with input \mbox{$w^t_e, e \in \Edgeset$}. On the other hand, if $\mu$ satisfies Assumption~1, $M(\mu)$ can also be computed by Algorithm~\ref{Matrix}.
%
%

    %
\begin{algorithm}[htb!]
    \DontPrintSemicolon
     \KwIn{$G_{m,n}$, $\tilde{w}_e, \forall e \in \Edgeset$.}
      Compute $H(u,v) := \sum \nolimits_{\Path \in \Pathset_{(u,v)}} {\prod \nolimits_{e \in \Path}{\tilde{w}_e} }$, $\forall u,v \in \mathcal{N}$. \;
      \For{$e_1=e_{[u_1,v_1]} \in \Edgeset$}{
      	$M(\mu_{\tilde{\boldsymbol{w}}})_{e_1, e_1} = \frac{H(s,u_1) \tilde{w}_{e_1} H(v_1,d)}{H(s,d)} $.\;
        \For{$e_2=e_{[u_2,v_2]} \in \Edgeset, e_2 > e_1$}{
        $M(\mu_{\tilde{\boldsymbol{w}}})_{e_1, e_2} = \frac{H(s,u_1) \tilde{w}_{e_1} H(v_1,u_2) \tilde{w}_{e_2}H(v_2,d)} {H(s,d)}$.
        }
       }
      \lFor{$e_1, e_2 \in \Edgeset, e_2\!<\!e_1$} {
      $M(\mu_{\tilde{\boldsymbol{w}}})_{e_1,e_2} \! =\! M(\mu_{\tilde{\boldsymbol{w}}})_{e_2,e_1}$.}
    \KwOut{The matrix $M(\mu_{\tilde{\boldsymbol{w}}})$.}
     \caption{Co-occurrence matrix computation.} \label{Matrix}
\end{algorithm}
Note that, we keep a generic notation in this algorithm: the input $\tilde{w}_e, e\in \Edgeset$ refers to any configuration of edges weights, not only those with the specific forms $w^t_e$ under the exponential updating rule. The output $M(\mu_{{\tilde{w}}})$ is the co-occurrence matrix corresponding to the distribution $\mu_{\tilde{\boldsymbol{w}}}$ that draws a path $\Path^*$ with probability 
    \begin{equation}
    \label{mu(w)}
        \mu_{\tilde{\boldsymbol{w}}}(\Path^*) = \prod \nolimits_{e\in \Path^*}{\tilde{w}_e} / \sum \nolimits_{\Path \in \Pathset}{(\prod \nolimits_{e\in \Path}{\tilde{w}_e})}.     
    \end{equation}
In Algorithm~\ref{Matrix}, we also drop the superscript $t$ in the notation of $H(u,v)$; these quantities can be efficiently computed (with inputs $\tilde{w}_e$) similar to $H^t(u,v)$ (with inputs $w^t_e$). The main intuition of  Algorithm~\ref{Matrix} is that if a path $\Path$ contains an edge \mbox{$e_1= e_{[u_1,v_1]}$}, then $\Path$ also has to contain a path from node $s$ to node $u_1$ and a path from node $v_1$ to node~${d}$. Similarly, if a path $\Path$ simultaneously contains the edges \mbox{$e_1= e_{[u_1,v_1]}$} and \mbox{$e_2 = e_{[u_2,v_2]}$}, then $\Path$ also contains a path from node $s$ to node $u_1$, a path from node $v_1$ to node $u_2$ and a path from node $v_2$ to node~$d$.
%

%
%
%
\subsection{$\EdAl$ - An Computationally Efficient Algorithm}
\label{sec:EdgeCB}
We now combine the techniques presented in the previous sections into a modified variant of $\ComAl$. This novel algorithm, denoted $\EdAl$, works on edges instead of paths. We also parameterize $\EdAl(\mu)$ with each corresponding exploration distribution $\mu$. Its pseudo code is given in Algorithm~\ref{AlgoEdge}.
 \begin{algorithm}[htb!]
    \DontPrintSemicolon
     \KwIn{$m,n,  T \in\mathbb{N}$,$\gamma \in [0,1], \eta>0$, distribution $\mu$.}
      $\forall e \in \Edgeset $, $w^{1}_e:= 1$.\;
     \For{$t=1,2,\ldots,T$}{
     	Adversaries play (unobserved by the learner).\;
        Sample $\beta$ from Bernoulli distribution $\mathcal{B}(\gamma)$.\;
        \lIf{$\beta=0$}{
        sample a path $\Path^t$ using the WP Algorithm with $\{w^{t}_e, e\in \Edgeset \}$.}
        \lElse{ Sample a path $\Path^t$ from distribution $\mu$.
        }
      	Suffer and observe the loss $L(\Path^t)\!=\!{\left(\Loss^t\right)\!}^{\top}\!{\Path}^t \le 1$.\;
        Compute $C^t:= \mathbb{E}_{\Path \sim d^t}[\Path \Path^{\top}]$ based on \eqref{eq:MatC} and Algorithm~\ref{Matrix}.\; 
        Estimate loss $\hat{\Loss}^t = \left(\Loss^t ({{\Path}^t})^{\top} \right) C^{-1}_t {\Path}^t$. \;
        $\forall e \in \Edgeset$, $w^{t+1}_{e}:= w^t_e \cdot e^{-\eta \hat{\Loss}_e^t }$.
        }
     \caption{ $\EdAl$($\mu$)~Algorithm for \textsc{PathCB}.} \label{AlgoEdge}
 \end{algorithm}
We conclude this section with the following~proposition.
 \begin{proposition}
 \label{propo:Ebandit}
    \emph{With the same choices of $\gamma$ and $\eta$, the expected regret of $\EdAl(\mu)$ is equal to that of $\ComAl(\mu)$ in \emph{PPPs}; thus, $\EdAl$ has the same regret bound as indicated in Theorem~\ref{BianchiTheo}. Given a distribution $\mu$ on $\Pathset$ that satisfies Assumption~1, $\EdAl(\mu)$ runs in $\mathcal{O}( n^2m^4 T) $; this is in contrast with $\ComAl(\mu)$ that runs in $\mathcal{O}( \exp(n) T)$.}
\end{proposition}

     %
\section{OPTIMIZING THE EXPLORATION DISTRIBUTION AND NUMERICAL EVALUATION}
\label{sec:Blackbox}
In this section, we investigate the exploration distribution $\mu$ that is used in both $\ComAl(\mu)$ and~$\EdAl(\mu)$. Recall the notation $\lambda^*[M]$ for the smallest non-zero eigenvalue of a matrix $M$. In Theorem \ref{BianchiTheo}, the regret bound is of order $\mathcal{O}(\lambda^*[M(\mu)]^{-1/2})$; therefore, to minimize this bound, we need to search for $\mu$ that maximizes $\lambda^*[M(\mu)]$. In several $\Cban$ problems (see \cite{cesa2012}), the \textit{uniform distribution} on the action set, denoted $\muUni$, was proven to yield an optimal choice to use in $\ComAl$. However, it is not the case for general PPPs and particularly for \textsc{PathCB}$(m,n)$: the eigenvalue $\lambda^*[M(\muUni)]$ may be of order ${\Omega}(P^{-1})$ which yields a regret upper-bound that is exponentially large in terms of the number of edges (an example can be found in \cite{cesa2012}). Nevertheless, in all previous works that apply the $\ComAl$ algorithm to PPPs, e.g. \cite{gyorgy2007} and~\cite{sakaue2018}, $\muUni$ is used. Moreover, since it requires that $\gamma \le 1$, the bound given in Theorem \ref{BianchiTheo} can only be obtained if \mbox{$T \!\ge\! \left[n \log(P) \right]/\left[ \left(\lambda^*[M(\mu)]\right)^2 \left( \frac{E}{n} \!+\! \frac{2}{\lambda^*[M(\mu)]} \right) \right]$} (parameters tuned by \cite{cesa2012}). If $\lambda^*[M(\mu)]$ is too small, $\ComAl(\mu)$ and $\EdAl(\mu)$ can only work with extremely large time horizon $T$, which is impractical. For these reasons, the choice of exploration distribution to use in these algorithms is crucial. 

Formally, let us label the paths in $\Pathset$ by \mbox{$\Path_1, \Path_2, \ldots, \Path_P$}, we consider an eigenvalue-optimization problem as follows (its search space is $P$-dimensional):
\begin{align}
\text{maximize} &\quad  \lambda^*\left[ \sum_{i=1}^P \nolimits{x_i \cdot \left[\Path_i \Path_i ^{\top} \right]}\right] \label{1formul1} \\
\text{subject to} 	&\quad\boldsymbol{x}\in [0,1]^P, \sum_{i=1}^P \nolimits {x_i} = 1.\label{1formul3}
\end{align}
It is suggested in \cite{cesa2012} that the problem \eqref{1formul1}-\eqref{1formul3} can be solved by casting it into a semi-definite programming problem (SDP). An explicit formulation of this SDP can be found in~Appendix~\ref{sec:SDP_form}. In principle, this SDP can be solved exactly to find a distribution $\mu$ that maximizes $\lambda^*[M(\mu)]$. However, in practice, this SDP formulation still cannot be solved efficiently due to the fact that the feasible set still has dimension $P$ and that it contains a constraint relating to a summation of $P$ terms. In our simulation, standard SDP solvers\footnote{CVXOPT solver, available at \textit{https://cvxopt.org/} and Mosek solver \textit{https://www.mosek.com/}, both use primal-dual interior points methods.} take a long running time to solve this SDP problem even with small instances and they easily run into computationally memory issues with moderate~instances.
\subsection{Derivative-free Optimization and Change of Search Space}
The challenge is to find a fast method that provides an exploration distribution $\mu$ to be used in $\EdAl(\mu)$ that guarantees a sufficiently good regret-bound. Moreover, it is desired to be able to efficiently sample a path from $\mu$ (line 6 in Algorithm~\ref{AlgoEdge}) and to efficiently compute the matrix $M({\mu})$ in order to compute $C^t$ (line 8 in Algorithm~\ref{AlgoEdge}). To do this, we reformulate the problem \eqref{1formul1}-\eqref{1formul3} to reduce the dimension of the search space. We consider the following problem whose search space is $E$-dimensional:
\begin{equation}
\label{eq:reduced}
 \max \limits_{{\boldsymbol{w} \in [0,\infty)^E}}{\lambda^*(M(\mu_{\tilde{\boldsymbol{w}}}))}.
\end{equation}
Here, we recall the notation $\mu_{\tilde{\boldsymbol{w}}}$---the distribution on the paths set (defined in~\eqref{mu(w)} for each $\tilde{\boldsymbol{w}} \in [0,\infty)^E$) such that each path weight is the multiplication of the corresponding edges weights. Therefore, for each feasible solution of~\eqref{eq:reduced}, say $\boldsymbol{w}^*$, we can construct a corresponding feasible solution $\mu_{\boldsymbol{w}^*}$ of~\eqref{1formul1}-\eqref{1formul3}; moreover, the objective function of~\eqref{eq:reduced} at $\boldsymbol{w}^*$ equals to that of~\eqref{1formul1}-\eqref{1formul3} at $\mu_{\boldsymbol{w}^*}$. The construction of $\mu_{\boldsymbol{w}^*}$ is in $\mathcal{O}(P)$ time, but we do not need to explicitly do so in order to run $\EdAl$ algorithm with $\mu_{\boldsymbol{w}^*}$. Instead, since $\mu_{\boldsymbol{w}^*}$ is guaranteed to satisfy Assumption~1, we can use the WP Algorithm to sample efficiently a path from $\mu_{\boldsymbol{w}^*}$ and use Algorithm~\ref{Matrix} to compute efficiently $M(\mu_{\boldsymbol{w}^*})$. Therefore, we can solve~\eqref{eq:reduced} to (implicitly) find an exploration distribution and use it efficiently in~$\EdAl$. 
\begin{figure}[htb!]
    \centering
    \includegraphics[height=0.12\textwidth]{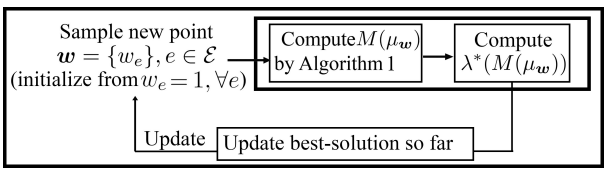}
    \caption{Diagram illustrating the derivative-free optimization.}
    \label{fig:Blackbox}
\end{figure}

Although~\eqref{eq:reduced} reduces significantly the dimension of the search space comparing to \eqref{1formul1}-\eqref{1formul3}, this formulation loses the structure that allows us to apply standard convex optimization algorithms.\footnote{The function giving the smallest non-zero eigenvalue of a matrix $M(\mu_{\boldsymbol{w}})$ from an input $\boldsymbol{w}$ is not known to be convex or concave.} Therefore, in this work, we use \emph{derivative-free} algorithms to heuristically solve~\eqref{eq:reduced}. Despite the fact that the solution found by this method may not be optimal, we can still guarantee that this solution is at least as good as the uniform distribution that is often used in the state-of-the-art algorithms (we initialize our algorithm with $\muUni$). Moreover, although the search space in~\eqref{eq:reduced} may not cover the whole search space in~\eqref{1formul1}-\eqref{1formul3}, the solution found in~\eqref{eq:reduced} (might be corresponding to a sub-optimal for~\eqref{1formul1}-\eqref{1formul3}) is guaranteed to be efficiently embedded with $\EdAl$; on the other hand, even if we found an optimal solution of~\eqref{1formul1}-\eqref{1formul3}, it does not guarantee to be efficiently used in $\EdAl$. A diagram explaining the intuition of our method to solve~\eqref{eq:reduced} can be found in Figure~\ref{fig:Blackbox}. We can use any derivative-free optimization solver that goes with specific strategies of sampling new points and justifying the current-best~solution.

We denote $\muFree$ the distribution corresponding to the solution of~\eqref{eq:reduced} found by our derivative-free method\footnote{Take $\boldsymbol{w}_e \!=\!1, \forall e \in \Edgeset$ (corresponding to $\muUni$) as the initialization~point.} and note that \mbox{$\lambda^* (M (\muFree)) \ge \lambda^* (M (\muUni))$}. Finally, as a corollary of Theorem~\ref{BianchiTheo} and Proposition~\ref{propo:Ebandit}, we have:
\begin{proposition}
\emph{In \textsc{PathCB}$(m,n)$, with appropriate parameters $\gamma$ and $\eta$, $\EdAl(\muFree)$ guarantees \mbox{$R_T\! \le\! \mathcal{O}\left( n   \sqrt{ \frac{2T}{ \lambda^*[M(\mu_\textrm{free})]}  } \right)$} and runs in $\mathcal{O}(n^2m^4 T)$~time.}
\end{proposition}

\subsection{Numerical Evaluation}
\label{evalua}
We conduct several experiments to evaluate the performance of  $\EdAl$ and measure the effect of optimizing the exploration distribution.\footnote{Our code is given at~\emph{https://github.com/dongquan11/BanditColonelBlotto}.} In these experiments, without loss of generality, a learner, having $m$ troops, plays a repeated $\CB$ game on $n$ battlefields with a single adversary who has $m_A$ troops. We define a special adversary, called the \emph{extreme-strong adversary}: an adversary having \mbox{$m_A\! = \!(n\!-\!1)(m\!+\!1)\! +\! (m\!-\!1)$} troops, she ``blocks'' $n\!-\!1$ battlefields (each has a value equal to $\varepsilon/(n-1)$) by allocating $m\!+\!1$ troops to them and allocating $m\!-\!1$ troops to a certain battlefield $i$ with value \mbox{$b_i\!=\!1\!-\!\varepsilon$} (unknown to the learner). 
In this case, the losses on all paths are always $1$ except for the single path representing that the learner allocates all $m$ troops to battlefield $i$; this path yields the loss $\varepsilon$. We choose this adversary to follow an example in \cite{cesa2012} illustrating why $\muUni$ fails to guarantee a good regret bound in PPPs. The algorithms need to ``explore" the low-loss path as soon as possible to reduce the~regret.

We use the \textsc{Zoopt} solver\footnote{Available at \textit{https://zoopt.readthedocs.io/en/latest/}. We run it in $100E$ iterations; this stopping criterion is recommended by \cite{hu2017}; moreover, this criterion is enough to solve~\eqref{eq:reduced} optimally in our experiments with small instances ($m,n \le 3$).} (see \cite{liu2017}) embedded with sRACOS algorithm (\cite{hu2017}) as the derivative-free optimization solver to  heuristically solve \eqref{eq:reduced} (its output is $\muFree$). Our experiments run on an Intel Core i5-7300U CPU@ 2.60GHz and 8.00GB RAM. Each instance is run 5 times and the average results are reported.

\begin{figure}[htb!]
    \centering
    { \begin{tikzpicture}
             \node (img1){{\includegraphics[height=0.16\textwidth]{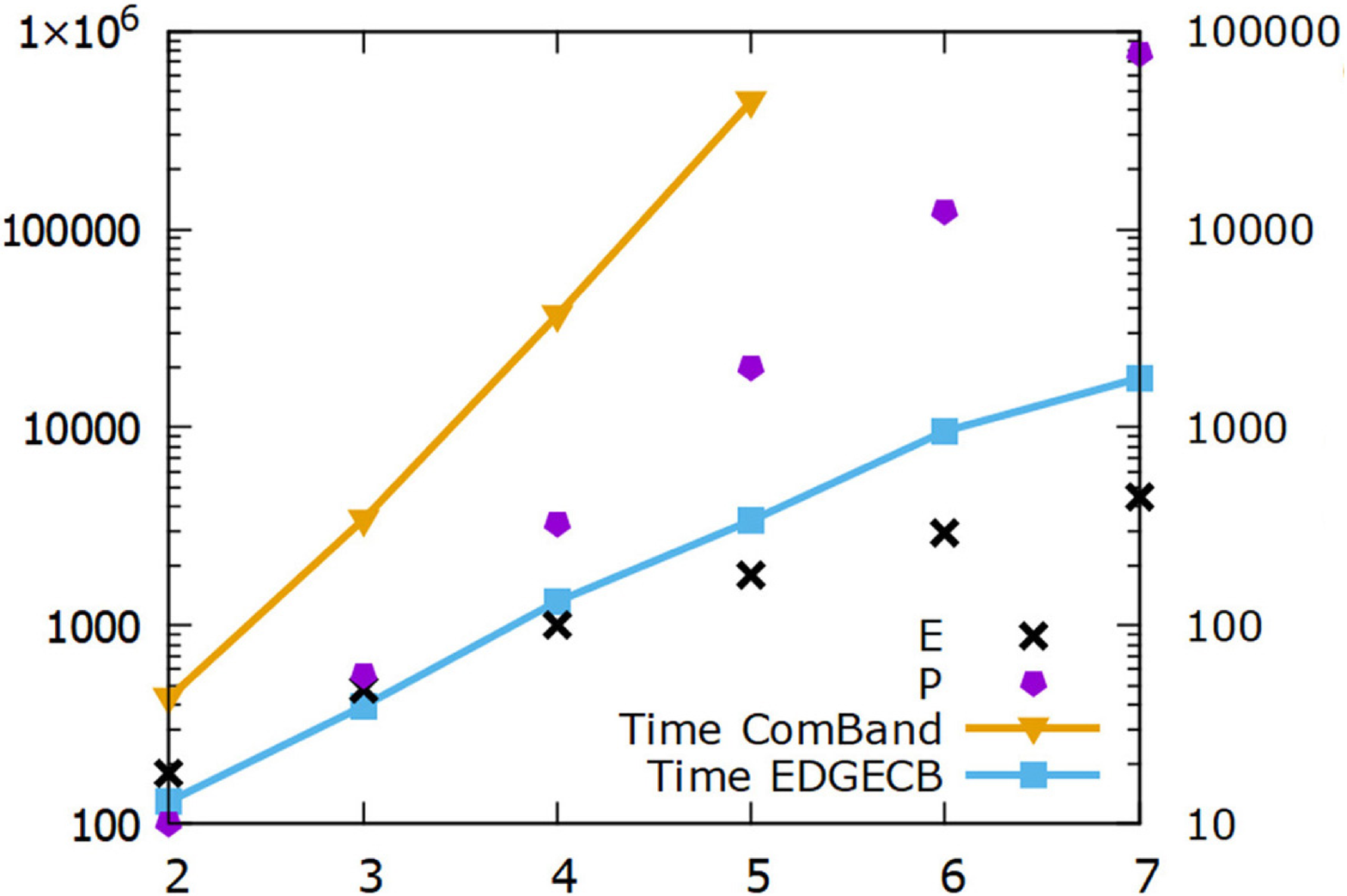} }};
             \node[below= 0.01cm of img1, node distance=0cm, yshift = 0.3cm] {\footnotesize $m$};
             \node[left= 0.01cm of img1, node distance=0cm, rotate=90, anchor = center] {\footnotesize Elapsed time (seconds)};
              \node[right= 0.01cm of img1, node distance=0cm, rotate=90, anchor = center] {\footnotesize Number of edges \& paths};
        \end{tikzpicture}
        }
    \caption{$\ComAl$($\mu_\text{free}$) vs $\EdAl$($\mu_\textrm{free}$); \mbox{$n=2m$}, $T=40000$ fixed.}
    \label{fig-first}
\end{figure}

In the first experiment, we compare the running time between $\ComAl$ and $\EdAl$ and the results confirm that $\ComAl$ takes exponential time while $\EdAl$ runs in polynomial time in terms of $m$ and~$n$; these results are reported in~Figure~\ref{fig-first} (the numbers of edges and paths in the corresponding $G_{m,n}$ are also reported for the sake of~comparison). 

\begin{figure}[htb!]%
       	\centering
        \subfloat[\mbox{$n\!=\!2m$}, $T\!=\!40000$ fixed.]{ \begin{tikzpicture}
             \node (img1){{\includegraphics[height=0.15\textwidth]{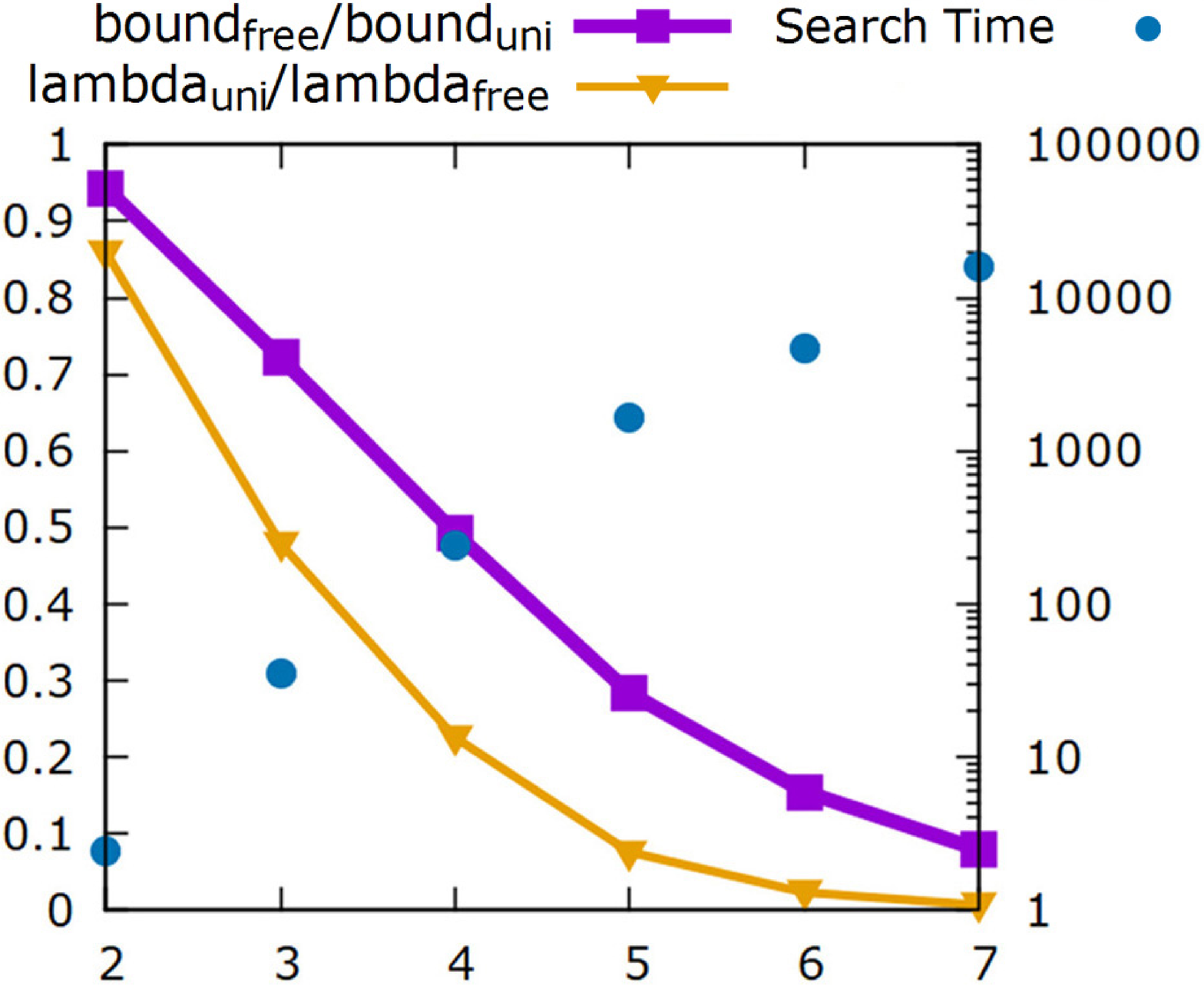} }};
             \node[below= 0.01cm of img1, node distance=0cm, yshift = 0.2cm] {\footnotesize $m$};
             \node[left= 0.01cm of img1, node distance=0cm, rotate=90, anchor = center] {\footnotesize Ratios};
              \node[right= 0.01cm of img1, node distance=0cm, rotate=90, anchor = center] {\footnotesize $\lambda_{\textrm{free}}$ Search Time};
        \end{tikzpicture}
        }%
        \subfloat[The actual regrets.]{
            	\begin{tikzpicture}
             \node (img1){	{\includegraphics[height=0.14\textwidth]{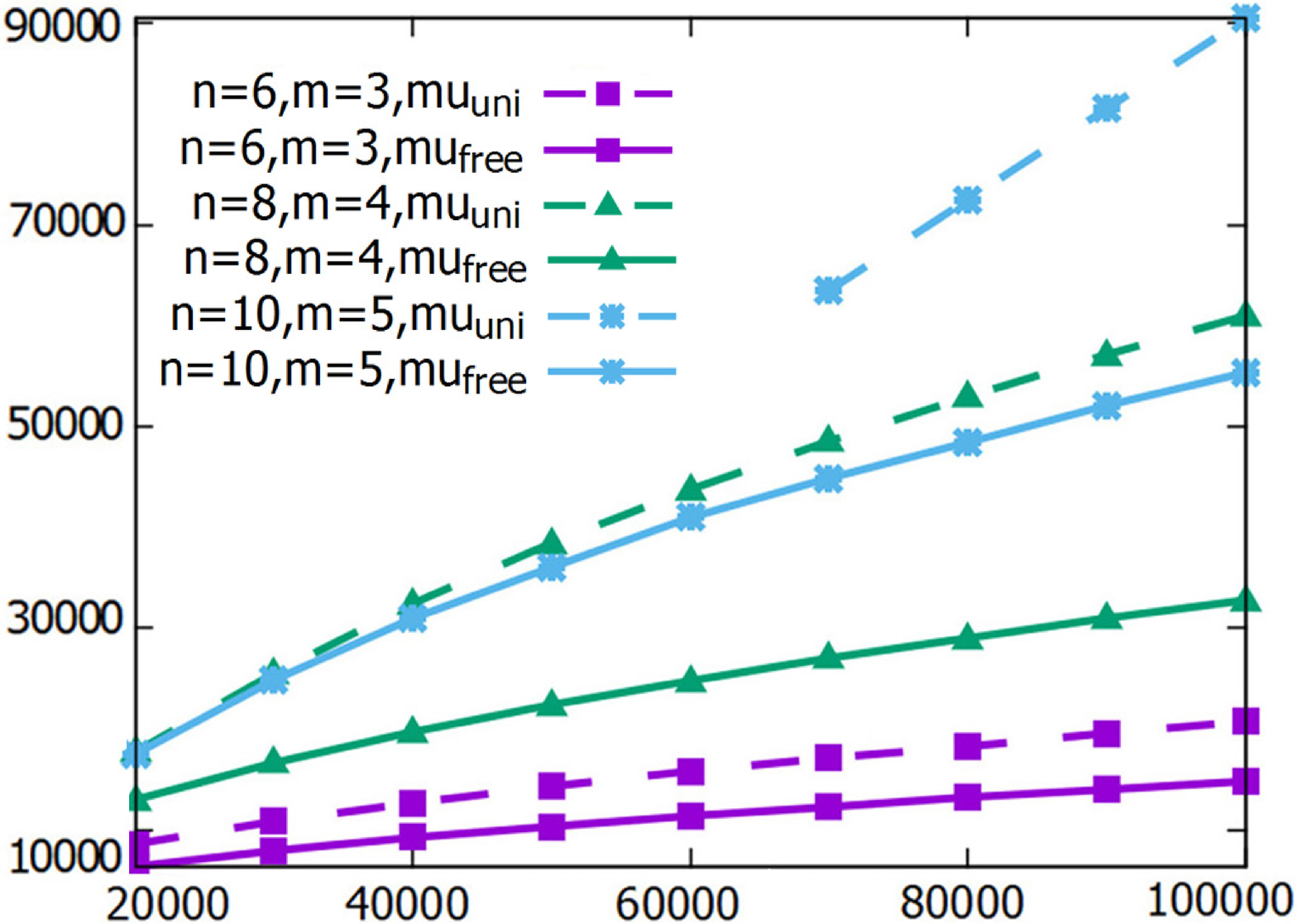} }};
             \node[below= 0.01cm of img1, node distance=0cm, yshift = 0.2cm] {\footnotesize $T$};
             \node[left= 0.01cm of img1, node distance=0cm, rotate=90, anchor = center] {\footnotesize Regret};
         \end{tikzpicture}
         }
        \caption{Performances evaluation of $\EdAl(\muUni)$ and $\EdAl(\muFree)$.}%
        \label{Figureall}%
      %
\end{figure}

Next, we compare the performances of $\EdAl$ when it uses $\muUni$ and $\muFree$ as the exploration distribution. Figure~\ref{Figureall}(a) ($y$-axes is drawn with log-scale) illustrates the trade-off between the time spending to find $\muFree$ and the improvement in the eigenvalues and the upper-bounds predicted by Theorem~\ref{BianchiTheo}. Note that the smaller the ratios $\textrm{bound}_{\textrm{free}} / \textrm{bound}_{\textrm{uni}}$ and $\lambda_{\textrm{uni}} / \lambda_{\textrm{free}}$ are, the more improvement that $\EdAl(\muFree)$ provides comparing to $\EdAl(\muUni)$. Finally, we compare the performance of $\EdAl(\muUni)$ and $\EdAl(\muFree$) by their actual regrets (see Figure~\ref{Figureall}(b)). Note that to efficiently compute the best hindsight loss (it is non-trivial), we apply a dynamic programming algorithm extracted from \cite{Vu18a} that finds the best response against a set of allocations of the~adversary. We observe that the actual regret of $\EdAl(\muFree)$ is better than $\EdAl({\muUni})$; as $m$ increases, the difference between these regrets also increases. For example, for instance $m=3$, \mbox{$n=6$} and $T=10^5$, the ratio $(\text{Regret}_\textrm{uni}- \text{Regret}_\textrm{free})/\textrm{Regret}_\textrm{uni}$ equals $28\%$ while this ratio of instance $m=5$, \mbox{$n=10$}, $T=10^5$ is $38\%$. Note that $\EdAl$($\muFree$) can run with larger instances (in $m,n$) but we choose not to report here since $\EdAl$($\muUni$) is unavailable in these instances (it requires extremely large~$T$). Besides the extreme-strong adversary, for this experiment, we also consider several other adversary's strategies (see Appendix~\ref{appen_experiments} for more details) and we notice that the results from these cases are similar to that of the extreme-strong adversary~case.


\section{Conclusion}
In this work, we present the $\EdAl$ algorithm for learning in the Colonel Blotto game that is modeled as a path planning problem. $\EdAl$ improves the regret guarantees compared to benchmark algorithm thanks to our proposed method finding an improved exploration distribution. Moreover, our algorithm is always efficiently implementable. This work not only extends the scope of application of the Colonel Blotto game in practice (even for large instances) but also can be applied to more general path planning problems.  

\bibliography{mybibfile}  

\clearpage
\newpage

\section{Appendix}

\subsection{SDP Formulation of the Exploration-Distribution Optimization Problem}
\label{sec:SDP_form}
To formulate the problem \eqref{1formul1}-\eqref{1formul3} into a SDP, we first observe that for any distribution $\mu$ such that the paths set $\Pathset$ is spanned by the support of $\mu$, the matrix \mbox{$M(\mu)$} always has a fixed number of zero eigenvalues (denoted $K$) and this number can be easily computed.\footnote{$Rank(M(\mu)) < E$ is the size of the largest linear independent subset of $\Pathset$, which is fixed and only depends on the structure of the layered graph~$G_{m,n}$. \textit{Rank-nullity} theorem implies that $K$ is also fixed. We can compute $K$ by computing rank of any particular matrix, say $M(\mu_\textrm{uni})$.} Therefore, the problem of maximizing $\lambda^*[M(\mu)]$ is equivalent to maximizing the sum of $K+1$ smallest eigenvalues of $M(\mu)$ which is formulated~as:
\begin{align}
	  \textrm{minimize } & \quad (K+1)s+ \textrm{Tr}(Z)  \label{2formul1}\\
 \textrm{subject to }
			 &\quad Z  \succeq 0 \label{2}\\
     & \quad Z + \sum_{i=1}^P\nolimits{x_i \cdot \Path_i \Path_i^{\top}} + s I_{E}  \succeq 0. \label{2formul2}
\end{align}
Here, $\boldsymbol{x} \in [0,1]^P$ and $r, s \in \mathbb{R}$, $Z \in \mathbb{M}_{E\times E}$ are the variables. $I_{E}$ is the identity matrix and the notation \mbox{$X \succeq 0$} indicates that the matrix $X$ is positive semi-definite. This is trivially deduced from the Linear Matrix Inequalities representation of the sum of $K+1$ largest eigenvalues of the matrix (see e.g., \cite{nesterov94,vandenberghe1996semidefinite}).

\subsection{Additional Numerical Experiments}
\label{appen_experiments}
\begin{figure}[htb!]%
        	\centering
    	 \subfloat[Against uniform-adversary ]{	\begin{tikzpicture}
             \node (img1){	{\includegraphics[height=0.15\textwidth]{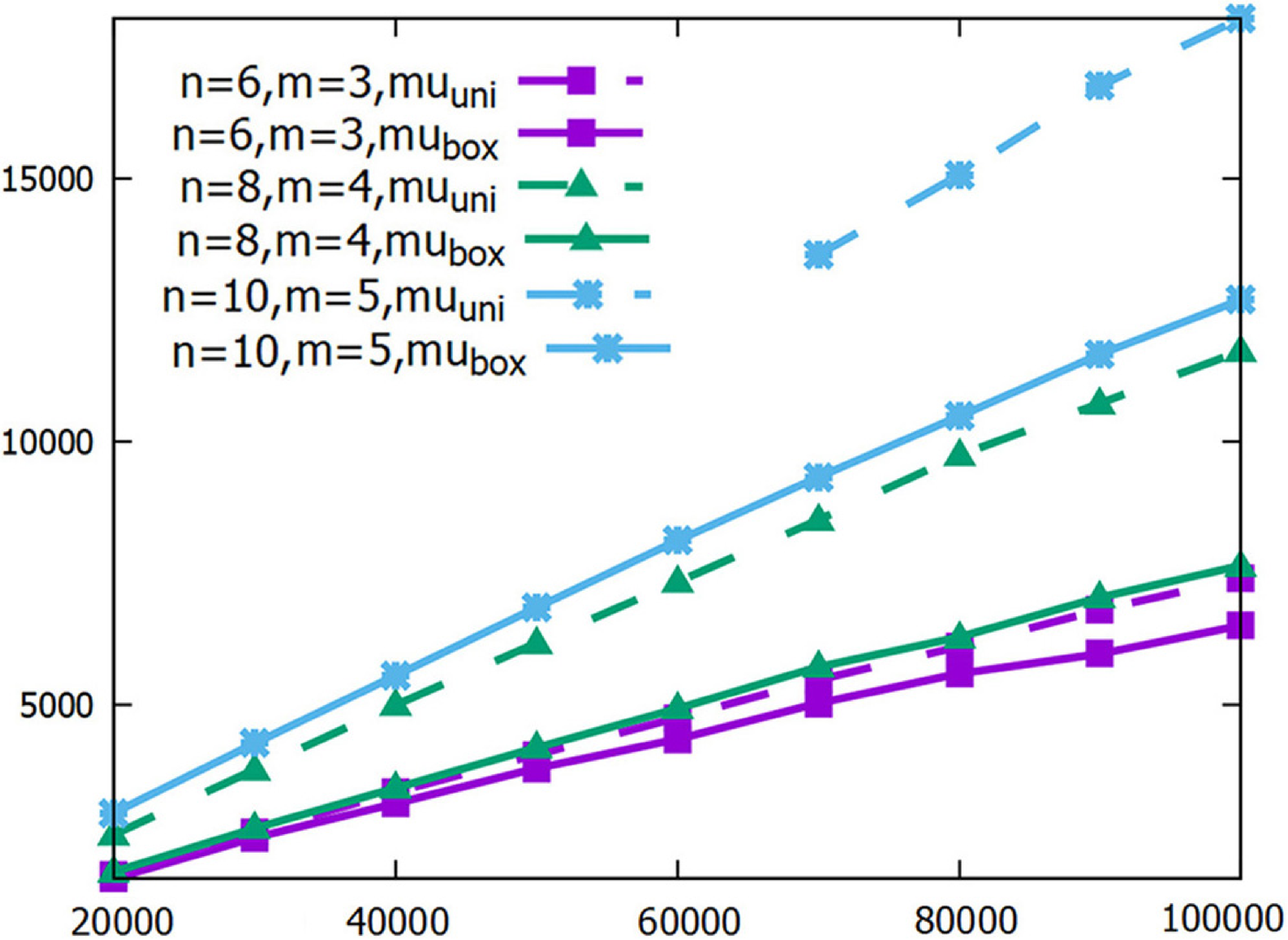} }};
             \node[below= 0.01cm of img1, node distance=0cm, yshift = 0.2cm] {\footnotesize $T$};
             \node[left= 0.01cm of img1, node distance=0cm, rotate=90, anchor = center] {\footnotesize Regret};
         \end{tikzpicture}
        }
        \subfloat[Against battlefield-wise adversary]{	\begin{tikzpicture}
             \node (img1){	{\includegraphics[height=0.15\textwidth]{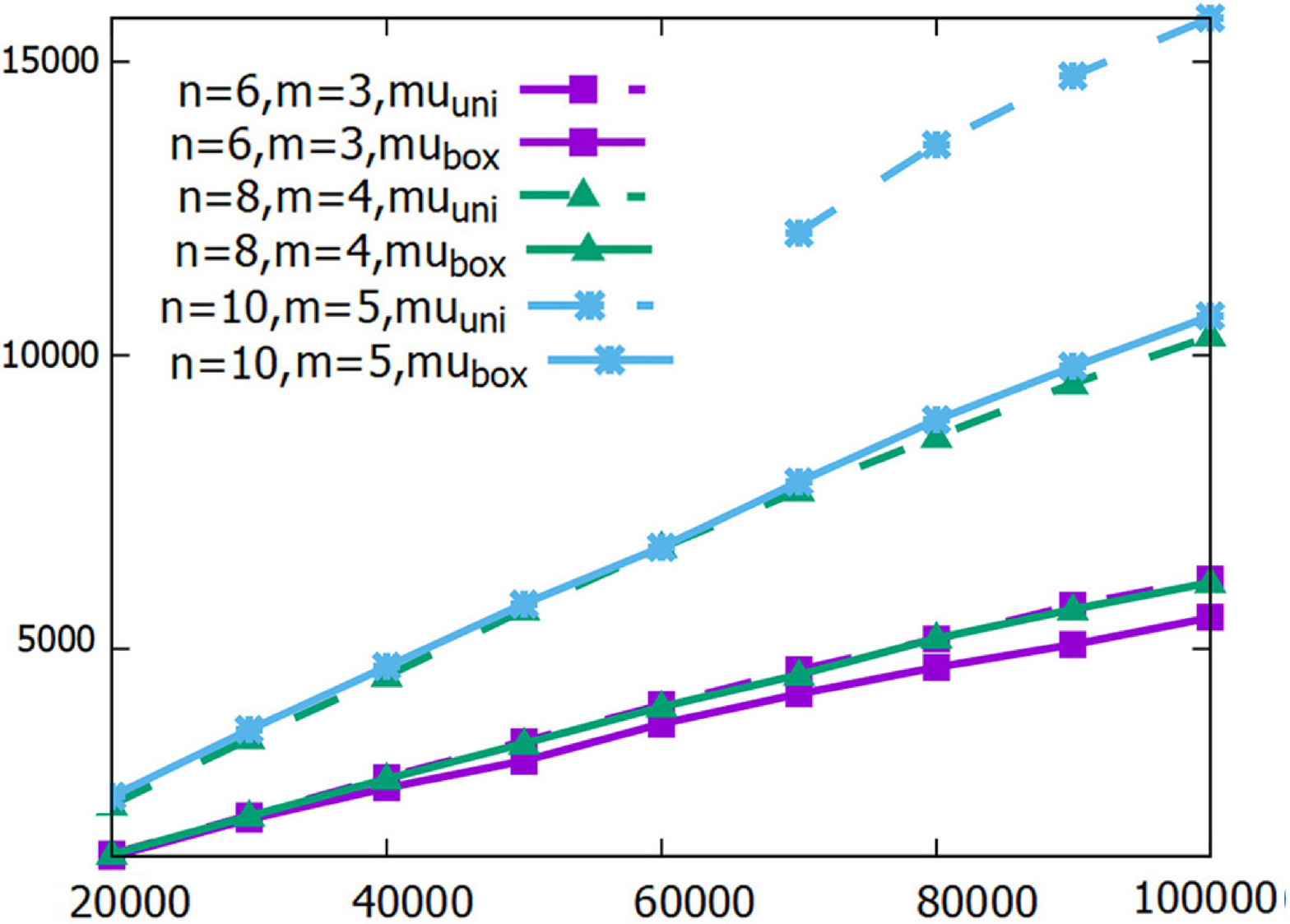} }};
             \node[below= 0.01cm of img1, node distance=0cm, yshift = 0.2cm] {\footnotesize $T$};
             \node[left= 0.01cm of img1, node distance=0cm, rotate=90, anchor = center] {\footnotesize Regret};
         \end{tikzpicture}
        }
           \caption{The actual regrets of $\EdAl(\muUni)$ and $\EdAl(\muFree)$.}
        \label{FigureActual}%
\end{figure}

 Besides the extreme-strong adversary, we also consider two other adversary's strategies: the \emph{uniform-adversary} (resp. the \emph{battlefields-wise adversary}) who at each time $t$ repeatedly draws a battlefield by uniform distribution (resp. draws battlefield $i$ with probability $b_i/ \sum \nolimits_{j \in [n]}{b_j}$) then allocates one troop to that battlefield until he runs out of troops ($m_{\mathcal{A}}\!=\!m$). For this experiment, the battlefields' values $b_i$ are generated uniformly from $[0,8]$. For each instance with different parameters $m,n$ and adversary's strategies, we run each algorithm \textsc{EdgeCB}$({\mu_{\textrm{uni}}})$ and \textsc{EdgeCB}($\mu_{\textrm{box}}$) $5$ times and the average results of their actual regret are reported in Figure~\ref{FigureActual}.

\end{document}